\documentclass[runningheads]{svmult}
\usepackage{makeidx}   
\usepackage{graphicx}  
\usepackage{subeqnar}  
\usepackage{multicol}  
\usepackage{cropmark} 
\usepackage{physprbb}  
\begin{document}
\title*{Large Scale Structure in an Era of Precision CMB Measurements}
\toctitle{CMB: II. Combined Opportunities in the Cosmic Microwave Background and the Large Scale Structure}
%
%
\titlerunning{CMB and LSS}
%
\author{Asantha Cooray}
\authorrunning{Asantha Cooray}
%
%

\institute{Theoretical Astrophysics Including Relativity Group,
California Institute of Technology, Pasadena, CA 91125, USA}

\maketitle              

\begin{abstract}
The advent of high signal-to-noise cosmic microwave background (CMB) anisotropy experiments
now allow detailed studies on the statistics related to temperature fluctuations. 
The existence of acoustic oscillations in the anisotropy power spectrum is now established with the clear detection of the first, and to a lesser confidence the second and the third, peak. Beyond the acoustic peak structure associated with CMB photon temperature fluctuations, we study the possibility for an observational detection  of oscillations in the large scale structure (LSS) matter power spectrum due to baryons. We  also suggest a new cosmological test using the angular power spectrum of dark matter halos, or clusters of galaxies detected via wide-field surveys of the large scale structure. The {\it standard  rulers} of the proposed test involve overall shape of the matter power spectrum and
baryon oscillation peaks in projection. The test allows a measurement of the angular diameter distance as a function of redshift, similar
to the distance to the last scattering surface from the first acoustic peak in the temperature anisotropy power spectrum.
The simultaneous detection of oscillations in both photons and baryons will provide a strong, and a necessary, confirmation of our 
understanding related to the physics during the recombination era. The proposed studies can be carried out with a combined analysis 
of CMB data from missions such as the MAP and the large scale structure data from missions such as the DUET.
\end{abstract}

\section{Oscillations in CMB}

The cosmic microwave background (CMB) is now a well known probe of the early universe.
The temperature fluctuations in the CMB, especially the so-called acoustic peaks in the angular power spectrum of CMB anisotropies, capture the physics of primordial photon-baryon fluid undergoing oscillations in the potential wells of the dark matter \cite{Huetal97}. The associated physics --- involving the evolution of a single
photon-baryon fluid under Compton scattering and gravity --- are both simple and linear, and
many aspects of it have been discussed in the literature since the early 1970s \cite{PeeYu70}. 
The gravitational redshift contribution at large angular scales \cite{SacWol67} and the photon-diffusion damping
at small angular scales \cite{Sil68} complete this description.
 
\begin{figure}[t]
\includegraphics[height=25pc,angle=-90]{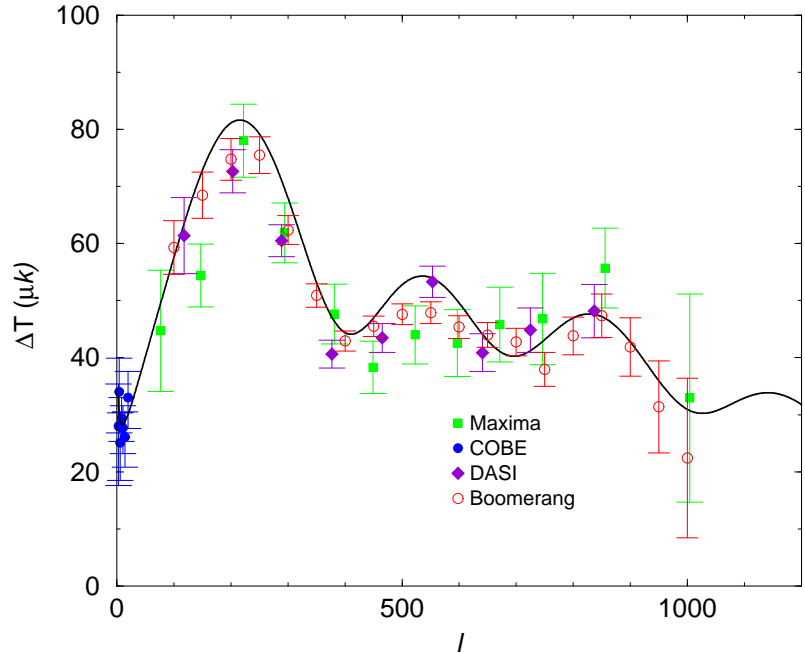}
\caption{Temperature fluctuations in the cosmic microwave background as seen by COBE (filled circles; \cite{Teg96}),
Boomerang (open circles; \cite{Netetal01}), MAXIMA (squares; \cite{Leeetal01}) and DASI (diamonds; \cite{Haletal01}).
The solid line shows the theoretical expectation for a cosmology with best-fit parameters for the Boomerang data following 
\cite{Netetal01}.}
\label{fig:cl}
\end{figure}

By now, there are at least five independent detections of the
first, and possibly the second and the third, acoustic peak
in the anisotropy power spectrum \cite{Miletal99,Haletal01}.  
We summarize these results in figure~\ref{fig:cl}.  Given the variety of experiments that are either collecting data or reducing  data that were recently collected, more detections that extend to higher peaks are soon expected. The NASA's MAP mission\footnote{http://map.gsfc.nasa.gov} is expected to provide a significant detection of the acoustic peak structure out to a multipole of $\sim$ 1000 and, in the long term,
the ESA's Planck surveyor\footnote{http://astro.estec.esa.nl/Planck/},
will extend this to a multipole of $\sim$ 2000 with better frequency coverage and polarization sensitivity. A discussion of these 
recent results and implications for cosmology are presented in the contribution to these proceedings by 
A. Melchiorri. 

An additional important aspect with respect to CMB is that in transit to us, photons are affected by the large scale structure along
the way. The modifications come from effects related to gravity, such as through frequency shifts associated with the integrated
Sachs-Wolfe effect \cite{SacWol67}, and through effects related to scattering such as the Sunyave-Zel'dovich (SZ) effect \cite{SunZel80}.
We refer the reader to Ref. \cite{Coo02} for a discussion of these contributions and their importance for understanding the
large scale structure via precision CMB measurements.

\begin{figure}[!h]
\includegraphics[height=23pc,width=16pc,angle=-90]{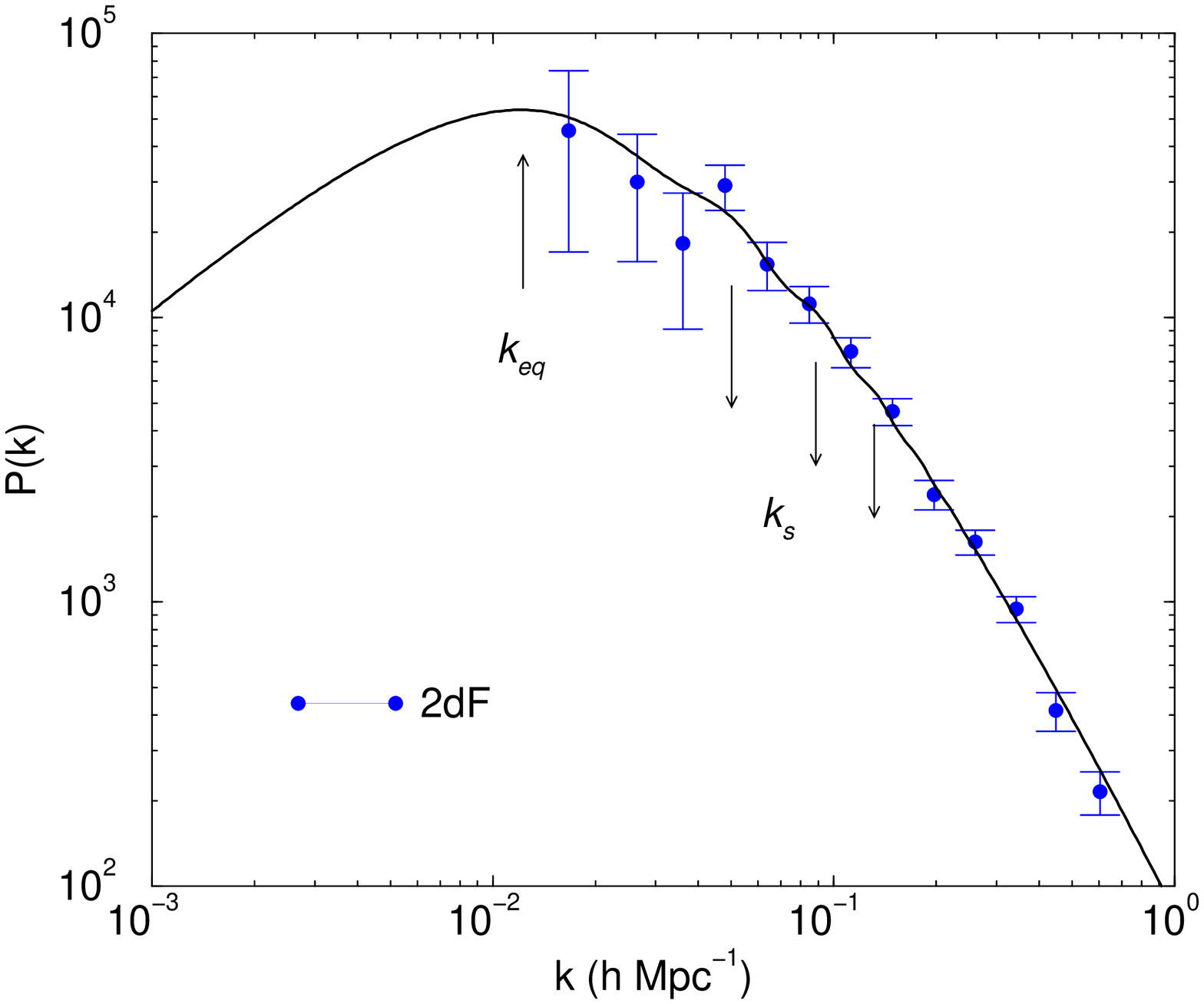}
\includegraphics[height=23pc,width=16pc,angle=-90]{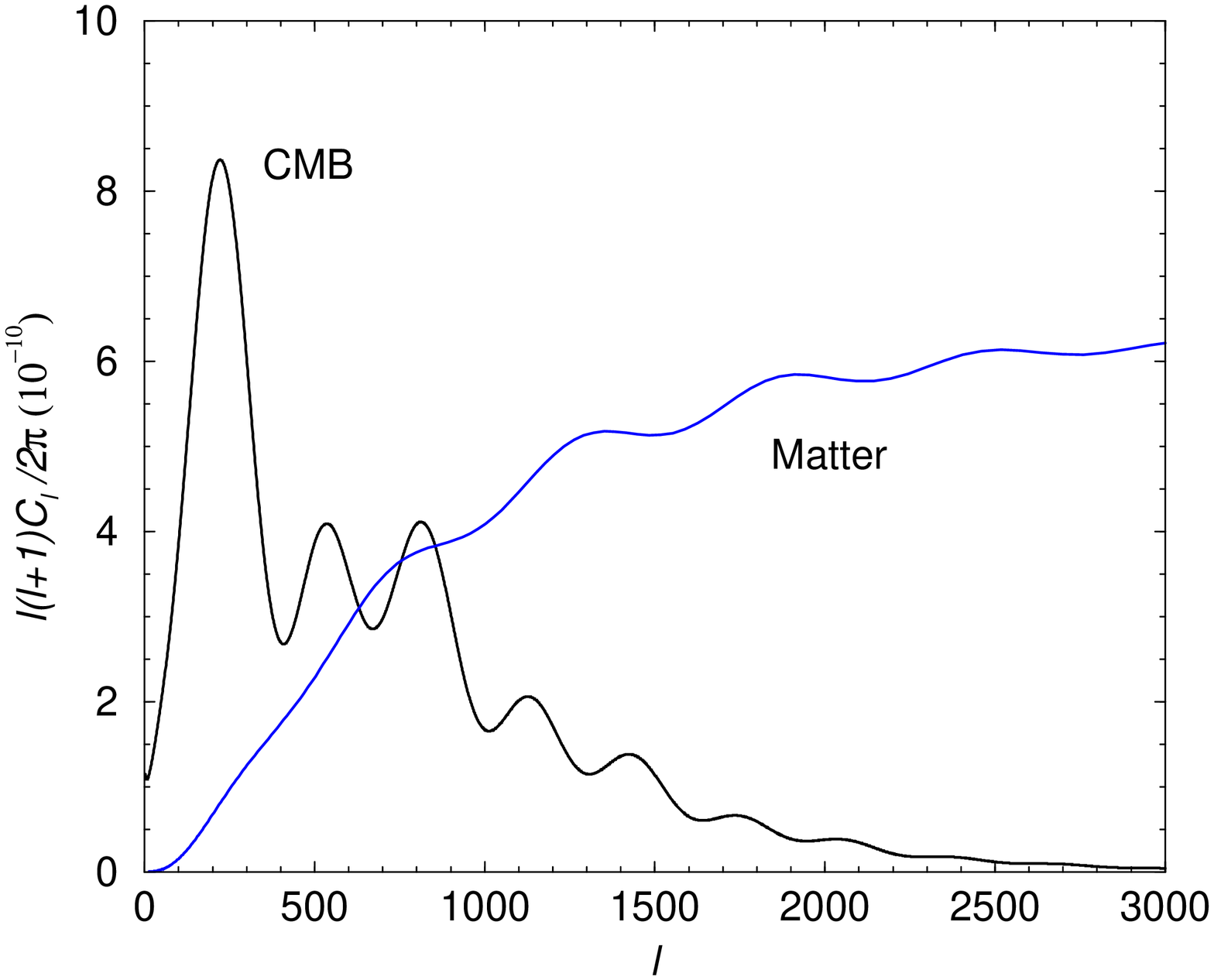}
\caption{{\it Top:} Large scale structure matter power spectrum. The data are from the analysis of 2dF redshift survey \cite{Tegetal99}.
We note two important physical scales of interest to the present discussion involving the matter-radiation equality $k_{\rm eq}$, which
produces a turnover in the power spectrum, and the peak location of baryon oscillations corresponding to the 
projection of the sound horizon at the end of the Compton-drag epoch $k_{\rm s}$. {\it Bottom:} A comparison of the oscillations in
CMB  due to photons and in matter due to baryons. We project the matter power spectrum at the distance of the last scattering surface
following equation~(\ref{eqn:cl}), and normalize it arbitrarily for illustrative purposes. 
Note that the acoustic oscillations in the matter power spectrum are out of phase with oscillations in the CMB anisotropy power
spectrum.}
\label{fig:pk}
\end{figure}

\section{Oscillations in LSS}

The detection of coherent oscillatory features in the CMB temperature anisotropy power spectrum 
suggests that if our physical description of them is correct, then, the large scale structure at low redshifts
should also contain a signature of oscillations. These troughs and peaks are associated with baryons which under went acoustic 
oscillations with photons as a single photon-baryon fluid  prior to recombination\footnote{The oscillations in CMB, due to radiation, 
and the oscillations in LSS, due to baryons, involve similar physics, though
they have subtle differences. These include a phase difference, $\pi/2$ at small wavenumbers,
 between the radiation and matter oscillatory peaks, due to a velocity term that source the growing mode of matter perturbations. Also, radiation oscillations project the sound horizon at the recombination while baryon oscillations project the sound horizon at the end of the Compton-drag epoch. 
In principle, these two epochs can be different, but for all purposes, the horizons are similar for currently
favorable $\Lambda$CDM cosmological models. We refer the interested reader to discussions in \cite{HuSug96,Meietal99}, 
and references therein, for further details.}.
Unlike oscillations in the angular power spectrum of CMB anisotropies, oscillations due to baryons in the 
matter power spectrum of the large scale structure are highly suppressed due to the low baryon content. They are also easily erased 
at late times due to the non-linear gravitational evolution of density perturbations \cite{Meietal99}. 
Compared to amplitudes of oscillations in the radiation, which vary by at least a factor of 2,
baryon oscillations have amplitudes which are significantly lower with
variations at the level of $\sim$ 8\% for currently favored $\Lambda$CDM cosmologies. These oscillations have
effective widths in Fourier space of order $\Delta k \sim 0.02$ h Mpc$^{-1}$.
As we discuss, these issues put several strong observational constraints for a reliable detection of these oscillations.

\begin{figure}[t]
\includegraphics[height=25pc,angle=-90]{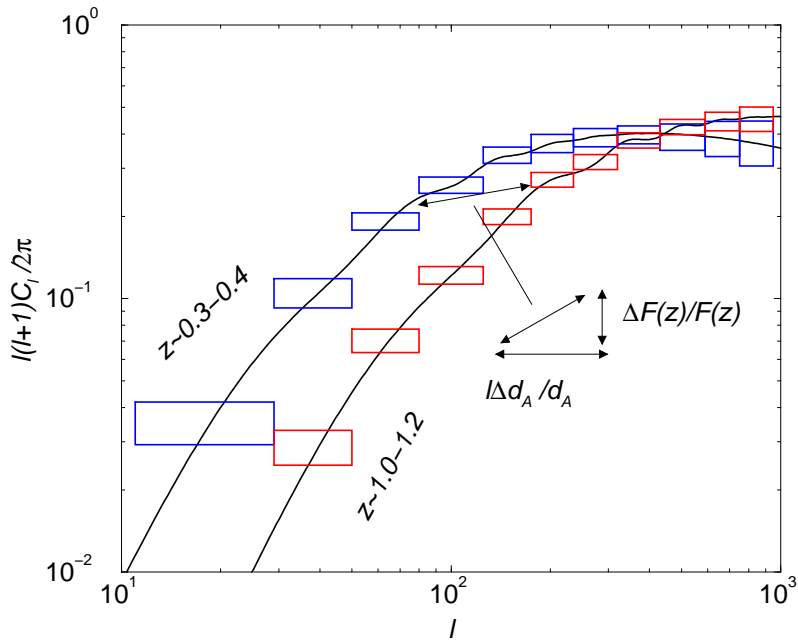}
\caption{Angular power spectrum of halos with $M> 10^{14} M_{\odot}$
in a wide-field survey
in bins of $z=0.3-0.4$ and $1.0-1.2$.  The binned
errors are 1-$\sigma$, and assume a survey of 4000 deg$^2$, within reach
of upcoming weak lensing and SZ surveys. The angular
power spectrum at high redshifts is shifted towards the right
proportional to the increase in the comoving angular diameter distance.}
\label{fig:halo}
\end{figure}

The verification on the presence of baryon oscillations in the large scale structure, 
however, is not impossible. Several attempts have already been made with three-dimensional
redshift surveys such as the 2dF Galaxy Redshift Survey and information from measurements related to angular clustering 
\cite{Peretal01},  though there is still no significant evidence for the presence of baryon oscillations.
The current three-dimensional surveys lack the required volume, with sizes of
order $L \sim 2\pi/\Delta k \sim 300$ h$^{-1}$ Mpc in all three dimensions, to reliably resolve the oscillations
while three dimensional power spectra obtained through an inversion of the angular correlation function should not
contain oscillatory features due to the nature of the inversion and the widths of window functions involved,
 given the expected widths of
oscillatory features \cite{Coo02a}.

While ongoing redshift surveys such as the Sloan Digital Sky Survey (SDSS; \cite{Yoretal00})
allow a strong possibility to detect these oscillations, in future, one can
also use projected clustering in Fourier space, 
such as the angular power spectrum similar to that of CMB anisotropies, to
detect baryon oscillations. 
This is useful given that one does not require precise redshift information but rather spatial information, and
estimates of redshift for binning purposes, from
wide-field surveys. In addition to galaxy survey data,
a number of additional observational efforts are under way or planned to image the large-scale structure of the universe out to redshift of a
few.  These wide-field surveys typically cover tens to thousands of square degrees on the sky and include the 
weak gravitational lensing shear observations with instruments such as the SNAP and the Large Aperture Synoptic Survey Telescope, 
dedicated small angular scale CMB telescopes that will map the SZ
effect through instruments such as the South Pole Telescope, and the wide-field 
X-ray imaging survey using the Dark Universe Explorer Telescope (DUET) mission. The latter was recently proposed
to NASA as a Medium Explorer mission.

In addition to their primary science goals, these surveys are
expected to produce catalogs of dark matter halos, which in the
case of lensing and SZ surveys, such catalogs are expected to be essentially
mass selected \cite{Witetal01}. In the case of X-ray, using cluster temperature data, one can
also construct a mass selected catalog of galaxy clusters. Lensing and other optical surveys are particularly promising in
that they will provide photometric redshifts on the member galaxies of a given halo;
this will render accurate determination of the halo redshift.  Halo number counts as a function of redshift
is a well-known cosmological test \cite{Haietal01}.  We can also consider the additional
information supplied by the angular clustering of halos, in particular, a possibility to use them for a detection of the
baryon oscillations, and to use clustering information for a new cosmological test.

\begin{figure}[t]
\includegraphics[height=25pc,angle=-90]{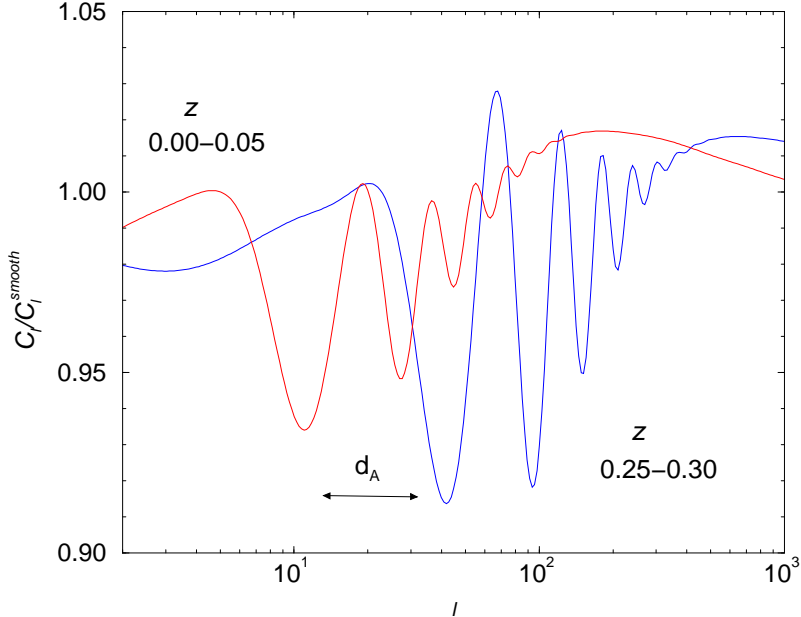}   
\caption{The normalized power spectrum of
halos at z $\sim$ 0.05 and z $\sim$ 0.3 measured in redshift bins of width $\Delta_z = 0.05$. 
The shift in oscillations is due to the projection and corresponds to the change in comoving angular diameter distance between the two redshift bins. The increase in amplitude of the oscillations in the higher redshift bin is due to the projection across a smaller width in the comoving angular diameter distance, though the bin width in redshift is same for the two curves.}
\label{fig:lssosc}
\end{figure}

It is well-known that a feature in the angular power spectrum of
known physical scale and originating from a known redshift can be used to measure the angular diameter
distance between us and this redshift; this has most notably
been applied to the case of CMB anisotropy power spectrum to
determine the distance to redshift $z \sim 10^3$. 
Here, one uses the peak location of the first acoustic peak which projects
on the multipolar space as $l_{\rm peak} = k_* d_A(z={\rm rec})$, where $k_*$ is the sound horizon at the recombination and the
$d_A$ is the associated angular diameter distance in comoving coordinates. Given the location,and a physical model for $k_*$,
one can constrain  cosmology
using the variation of $d_A$ as a function of cosmological parameters.

We can apply a similar argument to estimate cosmology using the clustering related to the large scale structure.
In figure~\ref{fig:pk}, we show the three-dimensional power spectrum and recent measurements from the 2dF galaxy
redshift survey, following the analysis by \cite{Tegetal99}. 
In the case of the large scale structure, there are two prominent features in the linear power spectrum associated with the
clustering of matter. These are the   horizon at the matter-radiation equality
\begin{equation}
k_{\rm eq} = \sqrt{2 \Omega_m H_0^2 (1+z_{\rm eq})} \propto \Omega_m h^2
\end{equation}
which controls the overall shape of the power spectrum, including a turn over, and the sound
horizon at the end of the Compton drag epoch, $k_{\rm s}(\Omega_m
h^2,\Omega_b h^2)$; the latter controls the peak location of  oscillations due to baryons in the power spectrum.
We denote these features in
figure~\ref{fig:pk}. 
Similar to the projection of the sound horizon at the recombination on to the CMB anisotropy power spectrum, 
the angular, or multipole, locations of features in the large scale power spectrum shift in
redshift as $l_{\rm eq, s} = k_{\rm eq, s} d_A(z_i)$. In \cite{Cooetal01}, this allowed us to 
propose the following test: measure angular power spectrum of projected clustering, $C_l^i$,
 in several redshift bins and, using the fact that $l_{\rm eq,s}$  scales with $d_A(z_i)$,
constrain the  angular diameter distance as a function of redshift.
Unlike the case with CMB, we can use tracers over a wide range in redshift
and measure the distance as a function of redshift, or redshift bin.
When combined with CMB observations, note that the absolute physical scale of $k_{\rm eq,s}$ can 
be directly calibrated as CMB anisotropies provide estimates of parameters such as $\Omega_mh^2$ and $\Omega_bh^2$ accurately. When combined with the large scale structure, one can break degeneracies, such as the one
between $\Omega_m$ and $h$, or given sufficient prior information on $h$ to constrain information
related to $\Omega_m$ and $\Omega_\Lambda$.

\begin{figure}[t]
\includegraphics[height=21pc,angle=-90]{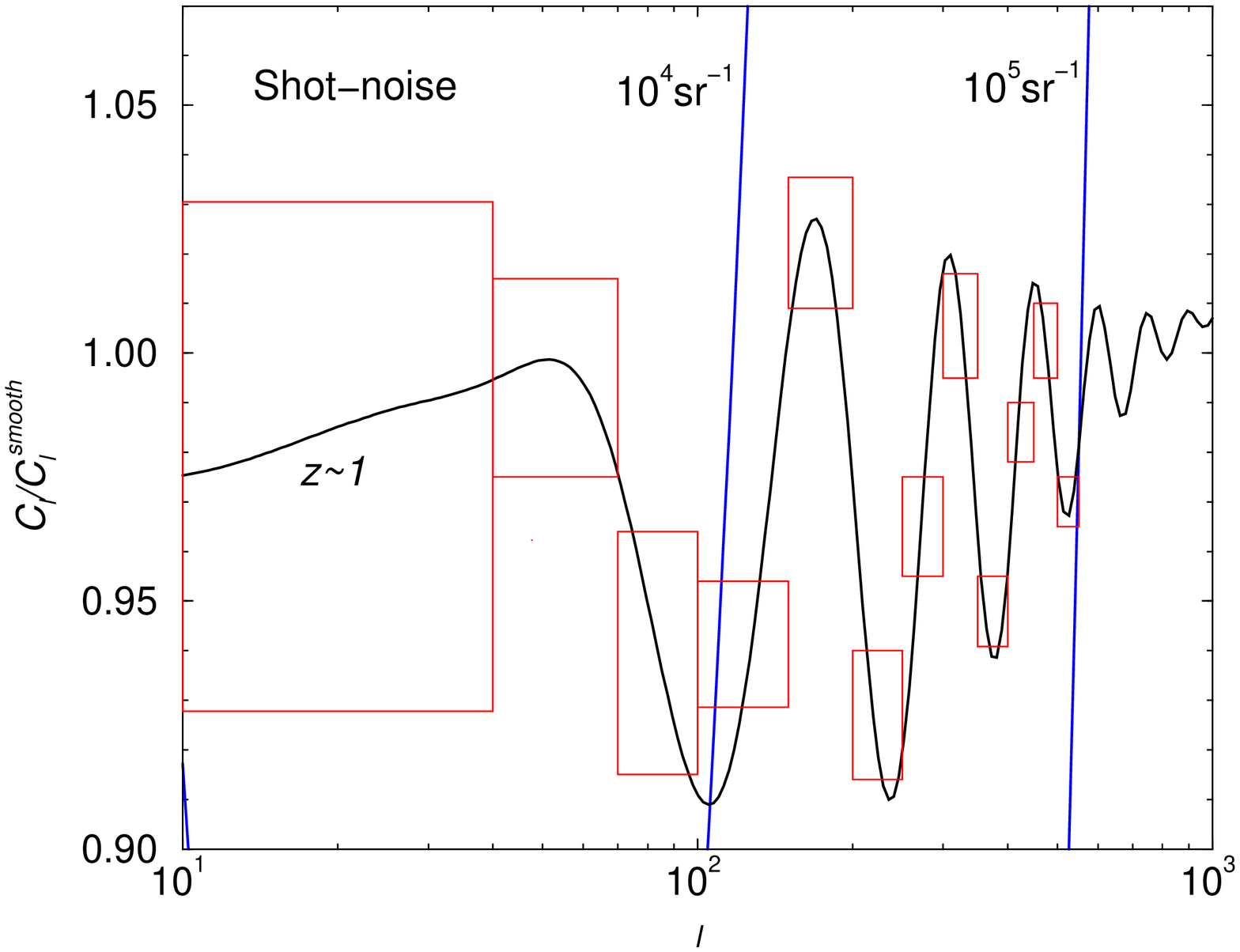}  
\includegraphics[height=21pc,angle=-90]{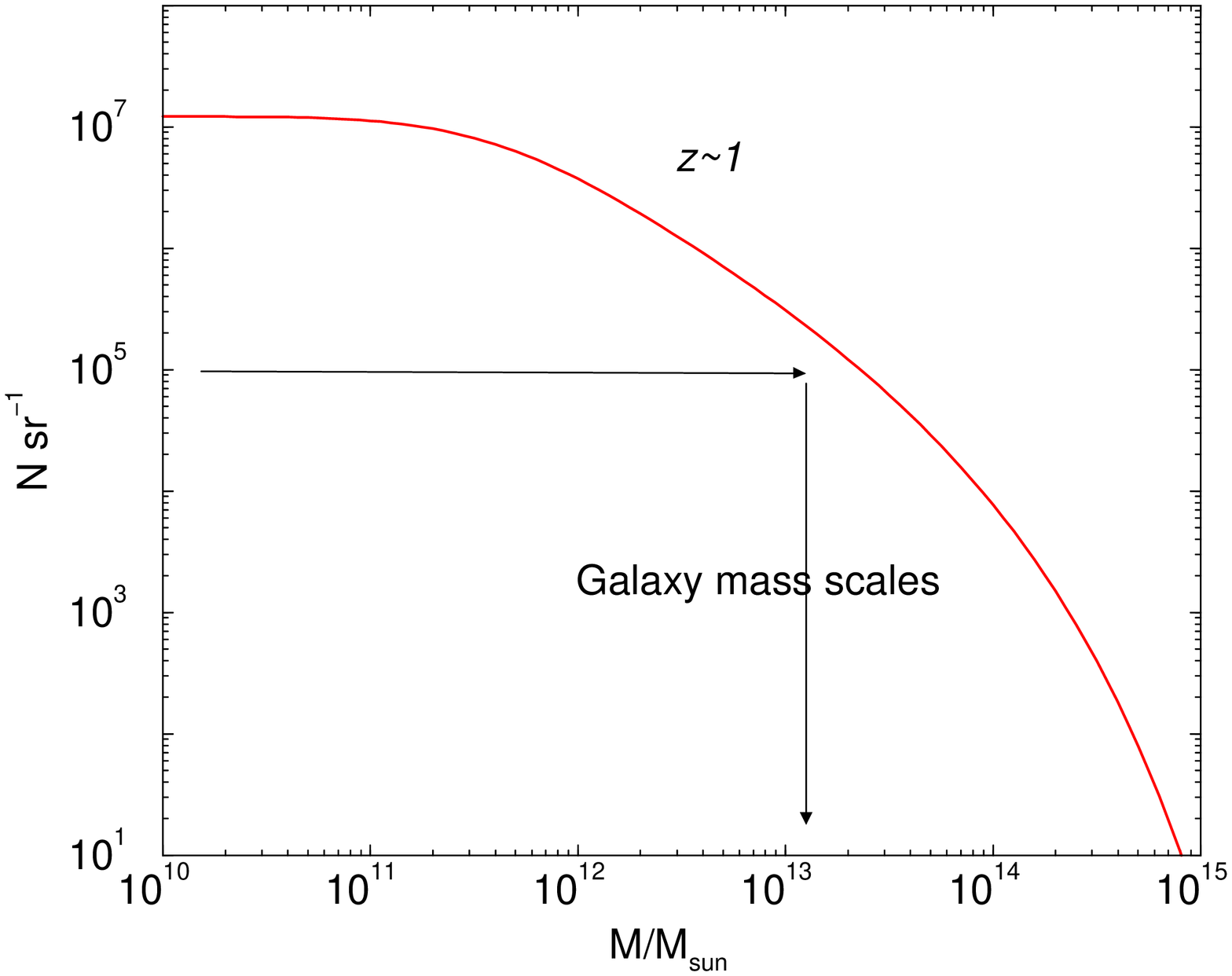}  
\caption{Expected cosmic variance associated with detection of oscillations around $z$ of 1, 
with a redshift bin of 0.1.  The shot-noise contribution due to the finite number of tracers are
shown, as a function of surface density of halos, with solid lines. 
In general, one requires a density of $10^{5}$ sr$^{-1}$ or more
per z-bin for a reliable detection. The  bottom panel shows the surface density of halos, as a function of halo
mass following the Press-Schechter theory \cite{PreSch74}. To obtain
a surface density of at least $10^5$ sr$^{-1}$, one needs to survey down to mass scales of $10^{13}$ M$_{\odot}$ at a
$z \sim 1$.}
\label{fig:nm}
\end{figure}

\begin{figure}[t]
\includegraphics[height=21pc,angle=-90]{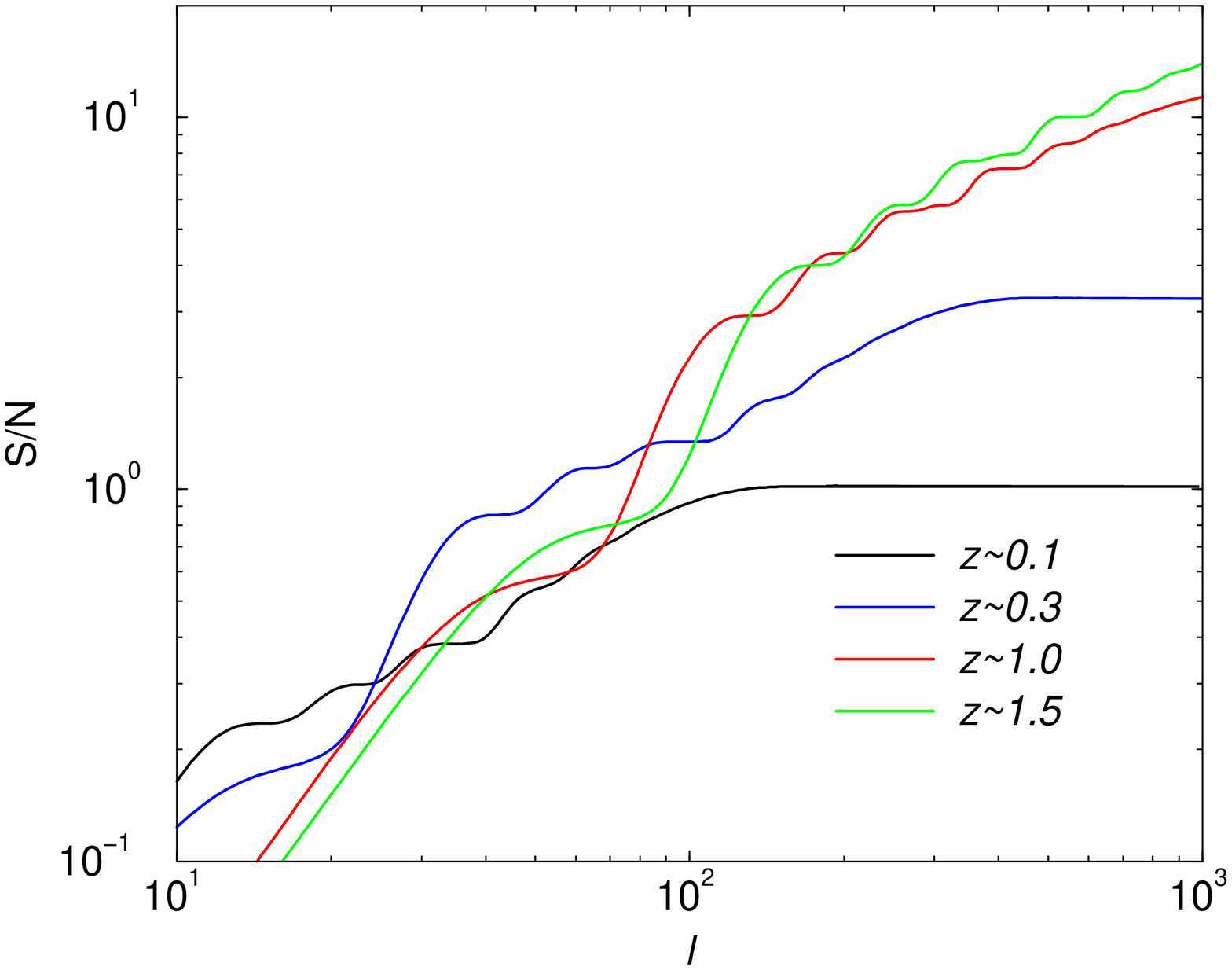}
\includegraphics[height=21pc,angle=-90]{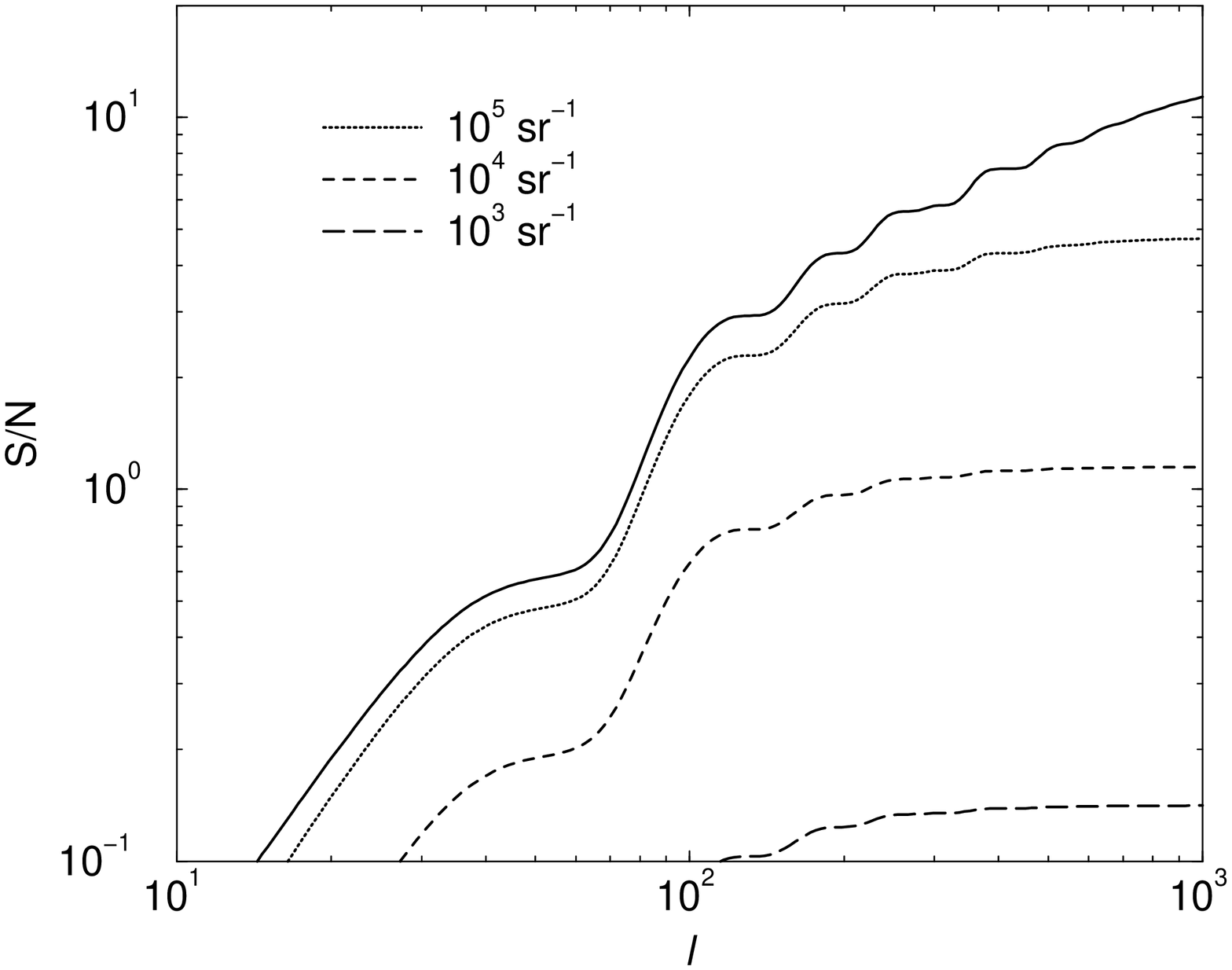}
\caption{The cumulative signal-to-noise ratios associated with the detection of baryons oscillations. The top
plot illustrates signal-to-noise values as a function of the mean redshift of bins with width 0.1. The bottom
panel shows the effect of finite number counts for the bin at $z \sim 1$. In general, one should expect to
 detect oscillations with a cumulative signal-to-noise ratio of $\sim$ 10.}
\label{fig:sn}
\end{figure}

To understand the projected clustering, consider the angular power spectrum of a tracer field of
large scale structure in a redshift bin. This is simply the
projection of the the tracer density power spectrum
\begin{equation}
C_l^i = \int dz\,W^2_i(z) {H(z) \over d_A^2(z)} P_{hh}\left(\frac{l}{d_A}; z\right) \,,
\label{eqn:cl}
\end{equation}
where $W_i(z)$ is the distribution of tracers in a
given redshift bin normalized so that
$\int\, dz\, W_i(z)=1$, $H(z)$ is the Hubble parameter.
For comparison, in the bottom panel of figure~\ref{fig:pk}, we show the projected  angular power spectrum of matter
at the last scattering surface and the angular power spectrum of CMB anisotropies. Note that the acoustic 
oscillations  associated with photons always show as peaks in the anisotropy power 
spectrum. There is an overall shift in oscillation phases in the 
projected angular power spectrum of matter at the last scattering surface 
when compared to those of CMB anisotropy power spectrum. 
If we have a tracer of the matter clustering  near recombination, say, just after the Compton-drag epoch, 
we should be able to make a comparison similar to the one suggested in the bottom panel of
figure~\ref{fig:pk}.

For the purpose  of this discussion, we assume a tracer of the large scale structure that correspond to 
mass selected halos, or galaxy clusters. If these halos trace the linear density field,
\begin{equation}
P_{hh}(k;z) = \left< b_M \right>^2(z) D^2(z) P^{\rm lin}(k;0)\,,
\end{equation}
where $\left<b_M\right>$ is the mass-averaged halo bias
parameter, $P^{\rm lin}(k;0)$ is
the present day matter power spectrum computed in linear theory, and $D(z)$
is the linear growth function
$\delta^{\rm lin}(k;z) = D(z)\delta^{\rm lin}(k;0)$.  A scale-independent halo bias
is commonly assumed in the so-called ``halo model'' \cite{CooShe02}
and should be valid at least in the linear regime.
Equation~(\ref{eqn:cl}) then becomes
\begin{eqnarray}
C_l^i & = & \int dz
 \,W_i^2(z) F(z) P^{\rm lin}\left(\frac{l}{d_A^i};0\right) \,, \\
\label{eqn:cz}
F(z) & = & {H(z) \over d_A^2(z)}  D(z)^2 \left< b_M \right>^2(z)\,.
\label{eqn:fz}
\end{eqnarray}
where the function $F_i(z)$, associated with redshift bin $i$,
contains information on the halo bias, the growth function, the
power spectrum normalization, and terms involved in the 3-D to
2-D projection (such as a $1/d_A^2$ term; see, Eq.~(\ref{eqn:cl})).
Note that $W_i(z)$ comes directly from the observations of the number counts as a function of redshift.         
 
\begin{figure}[t]
\includegraphics[height=21pc,angle=-90]{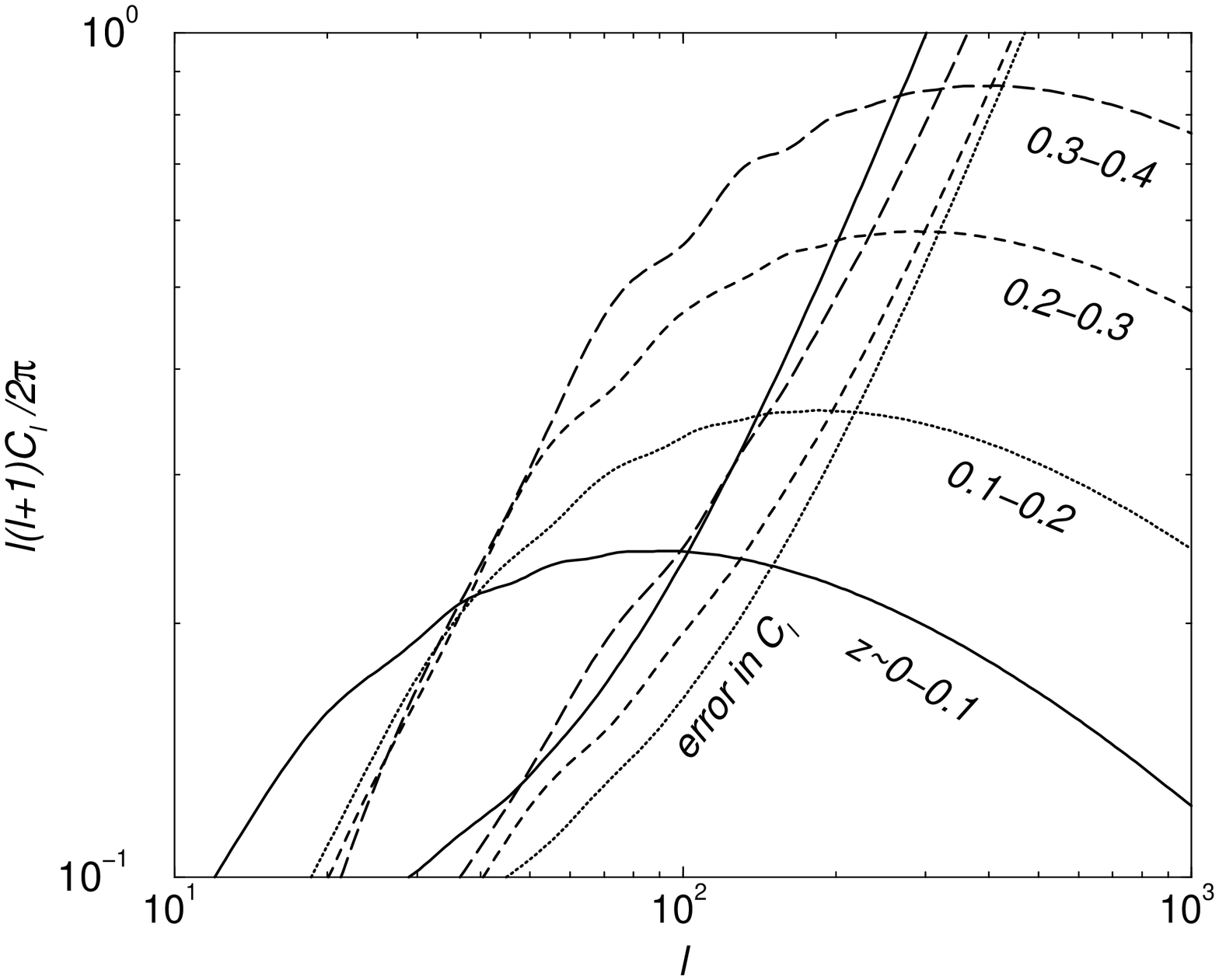}
\includegraphics[height=21pc,angle=-90]{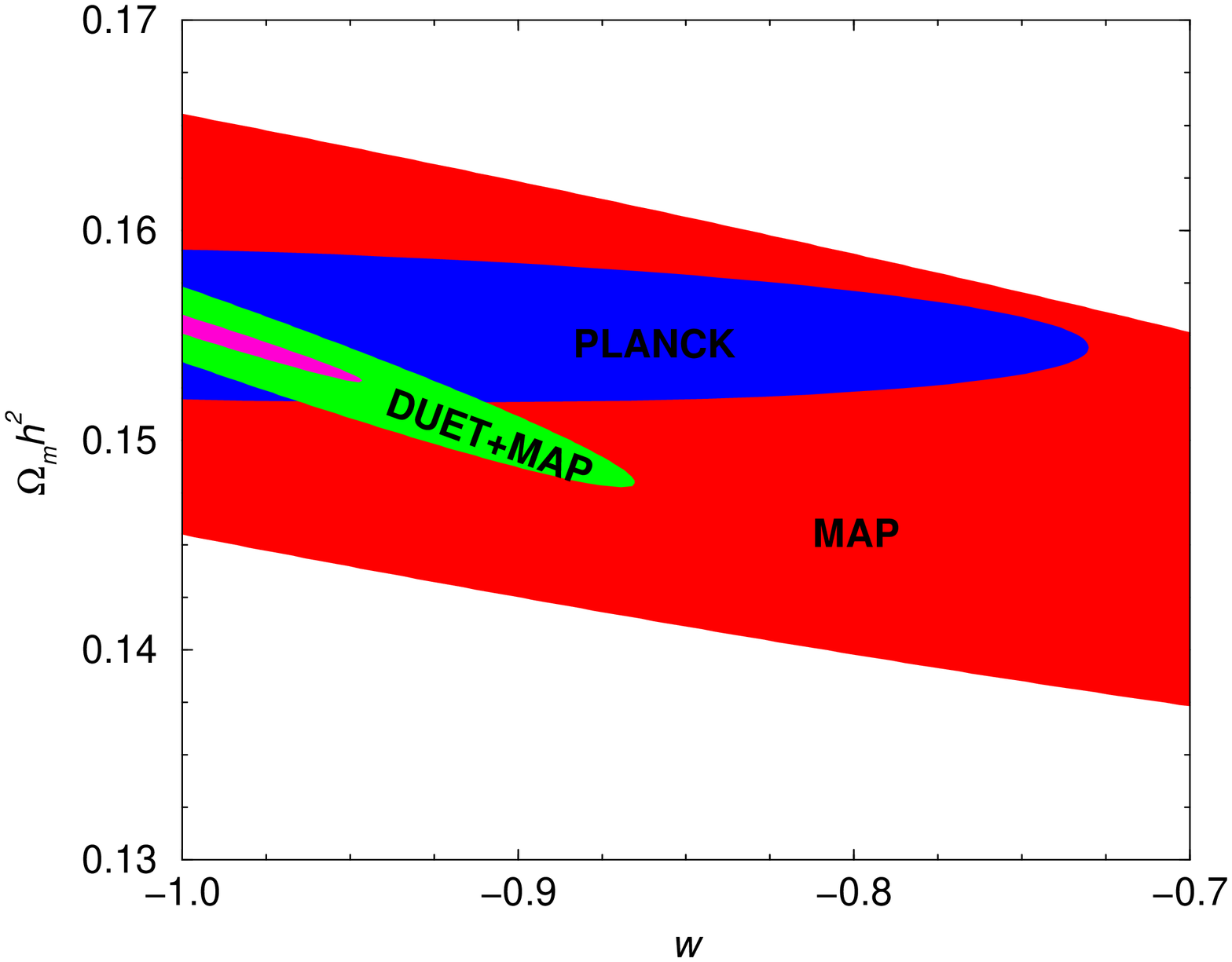}
\caption{The expected errors for detection of the clustering signal in catalogs produced by the proposed DUET mission.
As shown in the bottom panel, using the overall shape of the angular clustering power spectrum, 
due to the matter-radiation equality, and the projected locations of baryon oscillations, 
we can make significant measurements of cosmology. Combined with CMB data, such as the MAP and Planck missions, 
the DUET will
improve errors of certain parameters, such as properties related to the dark energy, by a factor of 5.}
\label{fig:duet}
\end{figure}

In figure~\ref{fig:halo}, we illustrate the proposed test.
The two curves show the halo power spectra in two redshift bins:
$0.3<z<0.4$ and $1.0<z< 1.2$. The angular power spectrum
corresponding to the higher redshift bin is shifted to the right
in accordance to the ratio of angular diameter distances $(\delta l/l
\sim \delta d_A/d_A)$. Much of this shift simply reflects the
Hubble law, $d_A \approx z/H_0$.  Since the physical size of the
two features --- the overall shape of the
spectrum and the baryon oscillations --- can be calibrated
from the CMB, these measurements can in principle
be used to determine the Hubble constant independently of the
distance ladder. In figure~\ref{fig:lssosc}, we remove the smooth curves and focus only on  oscillations.
It is clear from this figure that peak locations of oscillations shift in multipole space as
a function of the redshift, or more appropriately, the comoving distance.

The ability to measure oscillations crucially depend on both the cosmic variance and the shot-noise associated
with the tracer number counts. We illustrate this in figure~\ref{fig:nm}. In the top panel we show the
associated cosmic variance errors for an all-sky survey that image two-dimensional clustering at $z \sim 1$.
The vertical lines correspond to  errors resulting from the finite number of halos. For a reliable detection of the
first three baryon oscillations, one needs a survey with a tracer surface density of $10^5$ sr$^{-1}$ and this surface
density corresponds to a mass limit of $10^{13}$ M$_{\odot}$ at this redshift. The cumulative signal-to-noise
ratios are illustrated in figure~\ref{fig:sn}. These are the values to distinguish oscillations, which we
define for this purpose as any deviation about the mean prediction for clustering in the absence of any
baryons.

Note that in addition to the horizontal shift due to the change in angular diameter
distance, the power spectra in figure~\ref{fig:halo} are shifted
vertically due to the change in $F(z)$ (Eq.~\ref{eqn:fz}).
By ignoring the information contained in $F(z)$, the proposed geometric test in \cite{Cooetal01}
is robust against uncertainties in the mass selection, mass
function and linear bias.  Of course, if these uncertainties
are pinned down independently, both $F(z)$ and the halo abundance
in $W_i(z)$ will help measure the growth rate of structure. In \cite{Cooetal01}, we discussed in detail cosmological
information that can be gained from such a test using halos that will be detected in planned SZ and lensing
surveys.

We can perform a similar test with cluster catalogs that are expected to be produced with
upcoming wide-field  X-ray surveys such as the one planned with the proposed  DUET mission.
The DUET mission plans to map the $\pi$-steradians of the SDSS and the SZ deep fields
to  be observed with the South Pole Telescope. We show  expected errors and
improvements in cosmological parameters by combining the DUET clustering information with CMB data in figure~\ref{fig:duet}. The DUET mission allows a detection of baryon oscillations in the projected angular power
spectra of galaxy clusters at the few sigma level. This, by itself, is enough to break
significant degeneracies associated with cosmological parameter measurements from CMB data. 
Since the combined test measures cosmology through
the angular diameter distance at low redshifts, from clustered halos, and
at the last scattering, from the CMB acoustic peak, the test allows one to probe not
only the curvature but also relative abundances of various energy densities and properties of the dark energy.

The bottom panel of figure~\ref{fig:duet} shows errors on  $w$, the ratio of pressure to density of the
dark energy component, and $\Omega_mh^2$, from MAP and Planck, and improvements on these
errors when these CMB data sets are combined with clustering information from the DUET survey. There is at least 
a factor of 5 improvement in error associated with $w$. Such an improvement clearly demonstrate the need for combined studies
involving large scale structure and CMB. We strongly encourage combined approaches such as 
the one proposed with DUET and CMB data both to understand astrophysics, mainly
the presence of baryon oscillations, and cosmology, such as properties of the dark energy.

\section{Acknowledgments} We thank  Zoltan Haiman,  Wayne Hu and Dragan Huterer for collaborative work discussed in 
this article. The author is grateful to the DUET science team for providing him an opportunity to 
contribute aspects related to a combined CMB and large scale structure data analysis. 
The author acknowledges support from the Sherman Fairchild foundation and the Department of Energy and
 apologizes, before hand, for any missed references.

%


\begin{thebibliography}{8.}
\addcontentsline{toc}{section}{References}

\bibitem{Huetal97}
Hu W., Sugiyama N., Silk J. 1997, Nature 386, 37.

\bibitem{PeeYu70}
        Peebles, P.J.E. and Yu, J. T. 1970, ApJ, 162, 815;
	Sunyave, R. A. and Zel'dovich, Ya. B. 1970, Astrop. Space Sci., 7, 3.	

\bibitem{SacWol67}
Sachs, R. K., \& Wolfe, A. M., 1967, ApJ, 147, 73.

\bibitem{Sil68}
        Silk, J. 1968, ApJ, 151, 459.

\bibitem{Miletal99}
Miller, A. D. et al., 1999, ApJ, 524, L1;
        de Bernardis, P. et al., 2000, Nature, 404, 955;
Hanany, S. et al., 2000, ApJ, 545, L5.

\bibitem{Haletal01}
        Halverson, N. W. et al., 2001, preprint, astro-ph/0104489.

\bibitem{Teg96}
        Tegmark, M. 1996, ApJ, 464, 35.

\bibitem{Netetal01}
        Netterfield, C. B. et al. 2001, ApJ in press, astro-ph/0104460.

\bibitem{Leeetal01}
	Lee, A. T. et al. 2001, ApJ, 561, L1.

\bibitem{Tegetal99}
	Tegmark, M., Hamilton, A.J.S. and Xu, Y. 2001, MNRAS submitted (astro-ph/0111575).

\bibitem{SunZel80}
        Sunyaev, R.A. \& Zel'dovich, Ya. B. 1980, MNRAS, 190, 413.

\bibitem{Coo02}
	Cooray, A. 2002, in 2001 Coral Gables conference on cosmology and particle physics, Eds. B. Kursunoglu
       \& A. Perlmutter, American Institute of Physics Conference Proceedings (astro-ph/0203048).

\bibitem{HuSug96}
	Hu, W. and Sugiyama, N. 1996, ApJ, 471, 542.

\bibitem{Meietal99}
	Meiksin, A. White, M. and Peacock, J. A. 1996, MNRAS, 304, 851.

\bibitem{Peretal01}
	Percival, W. J. et al. 2001, MNRAS, 327, 1297;
	Miller, C. J., Nichol, R. C. and Batuski, D. J. 2001, Sci., 292, 2302.
	
\bibitem{Coo02a}
	Cooray, A. et al. 2002, in preparation.

\bibitem{Yoretal00}
	York, D. G. et al. 2000, AJ, 120, 1579.

\bibitem{Witetal01}
        Wittman, D., Tyson, J. A., Margoniner, V. E., Cohen, J. G.,
Dell'Antonio, I. P. 2001, ApJ, 557, L89;
	 Holder, G. P., Mohr, J. J., Carlstrom, et al. 2000, ApJ, 544, 629.  

\bibitem{Haietal01}
	 Haiman, Z., Mohr, J. J., Holder, G. P. 2001, 553, 545.

\bibitem{Cooetal01}
	Cooray, A., Hu, W., Huterer, D. and Joffre, M. 2001, ApJ, 557, L7.
	

\bibitem{PreSch74}
	 Press, W. H., Schechter, P. 1974, ApJ, 187, 425       

\bibitem{CooShe02}
        Cooray, A. \& Sheth, R. 2002, Physics Reports, submitted.



\end{thebibliography}
\end{document}